# Secure quantum dialogue via cavity QED

Tian-Yu Ye*

College of Information & Electronic Engineering, Zhejiang Gongshang University, Hangzhou 310018, P.R.China

**Abstract**

In this paper, a secure quantum dialogue protocol via cavity QED is suggested by using the evolution law of atom in cavity QED. The present protocol employs both the two-step transmission and the unitary operation encoding. Two security checks are adopted to ensure its transmission security against the active attacks from an outside eavesdropper. The present protocol avoids the information leakage problem by using the entanglement swapping between any two Bell states via cavity QED together with the shared secret Bell state. Compared with the previous information leakage resistant quantum dialogue protocol via cavity QED, the present protocol takes advantage in quantum measurement.

## 1 Introduction

As a particular kind of quantum secure direct communication (QSDC) [1-10], quantum dialogue aims to realize the bidirectional communication between two legitimate communicators. In 2004, Zhang *et al.*[11-12] and Nguyen[13] independently put forward this novel concept. Since then, many considerable achievements[14-22] have been accomplished with respect to this literature. However, in 2008, Gao *et al.*[23-24] and Tan *et al.*[25] independently pointed out that a security loophole, called 'information leakage' or 'classical correlation', always exists in quantum dialogue. Unfortunately, all of those previous quantum dialogue protocols[11-22] run this risk. As soon as it was discovered, the issue of information leakage in quantum dialogue quickly became the research focus in the literature of quantum secure communication. With the efforts of researchers, many quantum dialogue protocols without information leakage[26-33] came forth. In addition, in 2013, together with Jiang, the author[34] solved the information definite leakage problem in Man and Xia's protocol[16]. However, as it is pointed in Refs.[35-36],the two protocols in Ref.[34] still have the information leakage problem.

In this paper, a secure quantum dialogue protocol via cavity QED is suggested by using the evolution law of atom in cavity QED. The present protocol employs both the two-step transmission and the unitary operation encoding. Two security checks are adopted to ensure its transmission security against the active attacks from an outside eavesdropper. Moreover, the present protocol avoids the information leakage problem by using the entanglement swapping between any two Bell states via cavity QED together with the shared secret Bell state. Compared with the previous information leakage resistant quantum dialogue protocol via cavity QED, the present protocol takes advantage in quantum measurement.

## 2 Quantum dialogue protocol

Bell states are two-atom maximally entangled states, defined as

$$\left|\Phi^+\right\rangle = \frac{1}{\sqrt{2}}\left(\left|gg\right\rangle + \left|ee\right\rangle\right) = \frac{1}{\sqrt{2}}\left(\left|+\right\rangle\left|+\right\rangle + \left|-\right\rangle\left|-\right\rangle\right) \tag{1}$$

$$\left|\Phi^-\right\rangle = \frac{1}{\sqrt{2}}\left(\left|gg\right\rangle - \left|ee\right\rangle\right) = \frac{1}{\sqrt{2}}\left(\left|+\right\rangle\left|-\right\rangle + \left|-\right\rangle\left|+\right\rangle\right) \tag{2}$$

$$\left|\Psi^+\right\rangle = \frac{1}{\sqrt{2}}\left(\left|ge\right\rangle + \left|eg\right\rangle\right) = \frac{1}{\sqrt{2}}\left(\left|+\right\rangle\left|+\right\rangle - \left|-\right\rangle\left|-\right\rangle\right) \tag{3}$$

$$\left|\Psi^-\right\rangle = \frac{1}{\sqrt{2}}\left(\left|ge\right\rangle - \left|eg\right\rangle\right) = \frac{1}{\sqrt{2}}\left(\left|+\right\rangle\left|-\right\rangle - \left|-\right\rangle\left|+\right\rangle\right) \tag{4}$$

where $\left|\pm\right\rangle = \left(\left|g\right\rangle \pm \left|e\right\rangle\right)/\sqrt{2}$ . Four unitary operations are $U_{00} = I = \left|g\right\rangle\left\langle g\right| + \left|e\right\rangle\left\langle e\right|$ , $U_{01} = \sigma_x = \left|g\right\rangle\left\langle e\right| + \left|e\right\rangle\left\langle g\right|$ , $U_{10} = i\sigma_y = \left|g\right\rangle\left\langle e\right| - \left|e\right\rangle\left\langle g\right|$ and $U_{11} = \sigma_z = \left|g\right\rangle\left\langle g\right| - \left|e\right\rangle\left\langle e\right|$ . Let $U_{00}$ , $U_{01}$ , $U_{10}$ and $U_{11}$ denote bits 00, 01, 10 and 11, respectively.

Without loss of generality, suppose that both the atoms $A$ and $B$ and the atoms $C$ and $D$ are in the state of $\left|\Phi^+\right\rangle$. Two single-mode cavities, identical to those described in Refs.[21,32,37-40], are used for evolution of atoms. The atoms $A$ and $C$ are simultaneously sent into one single-mode cavity so that they simultaneously interact with it driven by a

E-mail：yetianyu@mail.zjgsu.edu.cn

classical field. In the meanwhile, the atoms $B$ and $D$ are simultaneously sent into the other single-mode cavity so that they also simultaneously interact with it driven by a classical field. The interaction time and Rabi frequency are ensured to satisfy $\lambda t = \pi/4$ and $\Omega t = \pi$ in both cases. Consequently, the total system will finally evolve into

$$\left|\Phi^+\right\rangle_{AB} \otimes \left|\Phi^+\right\rangle_{CD} = \frac{1}{2}\left(-i\left|gg\right\rangle_{AC}\left|ee\right\rangle_{BD} - i\left|ee\right\rangle_{AC}\left|gg\right\rangle_{BD} - i\left|ge\right\rangle_{AC}\left|eg\right\rangle_{BD} - i\left|eg\right\rangle_{AC}\left|ge\right\rangle_{BD}\right) \quad (5)$$

Extending $\left|\Phi^+\right\rangle_{CD}$ to other three Bell states $\left|\Phi^-\right\rangle_{CD}$, $\left|\Psi^+\right\rangle_{CD}$ and $\left|\Psi^-\right\rangle_{CD}$, if the evolving condition is the same as the above, the total system will respectively evolve into

$$\left|\Phi^+\right\rangle_{AB} \otimes \left|\Phi^-\right\rangle_{CD} = \frac{1}{2}\left(\left|gg\right\rangle_{AC}\left|gg\right\rangle_{BD} - \left|ee\right\rangle_{AC}\left|ee\right\rangle_{BD} - \left|ge\right\rangle_{AC}\left|ge\right\rangle_{BD} + \left|eg\right\rangle_{AC}\left|eg\right\rangle_{BD}\right) \quad (6)$$

$$\left|\Phi^+\right\rangle_{AB} \otimes \left|\Psi^+\right\rangle_{CD} = \frac{1}{2}\left(-i\left|gg\right\rangle_{AC}\left|eg\right\rangle_{BD} - i\left|ee\right\rangle_{AC}\left|ge\right\rangle_{BD} - i\left|ge\right\rangle_{AC}\left|ee\right\rangle_{BD} - i\left|eg\right\rangle_{AC}\left|gg\right\rangle_{BD}\right) \quad (7)$$

$$\left|\Phi^+\right\rangle_{AB} \otimes \left|\Psi^-\right\rangle_{CD} = \frac{1}{2}\left(\left|gg\right\rangle_{AC}\left|ge\right\rangle_{BD} - \left|ee\right\rangle_{AC}\left|eg\right\rangle_{BD} - \left|ge\right\rangle_{AC}\left|gg\right\rangle_{BD} + \left|eg\right\rangle_{AC}\left|ee\right\rangle_{BD}\right) \quad (8)$$

According to Eqs.(5)-(8), each result combination of the atoms $A$ and $C$ and the atoms $B$ and $D$ does correspond only to one initial state of the atoms $A$ and $B$ and the atoms $C$ and $D$. Four collections composed of different result combinations of the atoms $A$ and $C$ and the atoms $B$ and $D$ can be labeled as:

$$\left\{\left|gg\right\rangle_{AC}\left|ee\right\rangle_{BD}, \left|ee\right\rangle_{AC}\left|gg\right\rangle_{BD}, \left|ge\right\rangle_{AC}\left|eg\right\rangle_{BD}, \left|eg\right\rangle_{AC}\left|ge\right\rangle_{BD}\right\} \to C_0 \quad (9)$$

$$\left\{\left|gg\right\rangle_{AC}\left|gg\right\rangle_{BD}, \left|ee\right\rangle_{AC}\left|ee\right\rangle_{BD}, \left|ge\right\rangle_{AC}\left|ge\right\rangle_{BD}, \left|eg\right\rangle_{AC}\left|eg\right\rangle_{BD}\right\} \to C_1 \quad (10)$$

$$\left\{\left|gg\right\rangle_{AC}\left|eg\right\rangle_{BD}, \left|ee\right\rangle_{AC}\left|ge\right\rangle_{BD}, \left|ge\right\rangle_{AC}\left|ee\right\rangle_{BD}, \left|eg\right\rangle_{AC}\left|gg\right\rangle_{BD}\right\} \to C_2 \quad (11)$$

$$\left\{\left|gg\right\rangle_{AC}\left|ge\right\rangle_{BD}, \left|ee\right\rangle_{AC}\left|eg\right\rangle_{BD}, \left|ge\right\rangle_{AC}\left|gg\right\rangle_{BD}, \left|eg\right\rangle_{AC}\left|ee\right\rangle_{BD}\right\} \to C_3 \quad (12)$$

Further extending $\left|\Phi^+\right\rangle_{AB}$ to other three Bell states $\left|\Phi^-\right\rangle_{AB}$, $\left|\Psi^+\right\rangle_{AB}$ and $\left|\Psi^-\right\rangle_{AB}$, all the result collections composed of different result combinations of the atoms $A$ and $C$ and the atoms $B$ and $D$ from different initial states of the atoms $A$ and $B$ and the atoms $C$ and $D$ can be summarized in Table 1. From Table 1, it is easy to find out that the result collection derived from the entanglement swapping between any two initial Bell states via cavity QED is always solely located in one of the four result collections including $C_0$, $C_1$, $C_2$ and $C_3$.

Table 1. The result collections composed of different result combinations of the atoms $A$ and $C$ and the atoms $B$ and $D$ from different initial states of the atoms $A$ and $B$ and the atoms $C$ and $D$

| | $\left\|\Psi^+\right\rangle_{CD}$ | $\left\|\Psi^-\right\rangle_{CD}$ | $\left\|\Phi^+\right\rangle_{CD}$ | $\left\|\Phi^-\right\rangle_{CD}$ |
|---|---|---|---|---|
| $\left\|\Psi^+\right\rangle_{AB}$ | $C_1$ | $C_0$ | $C_3$ | $C_2$ |
| $\left\|\Psi^-\right\rangle_{AB}$ | $C_0$ | $C_1$ | $C_2$ | $C_3$ |
| $\left\|\Phi^+\right\rangle_{AB}$ | $C_2$ | $C_3$ | $C_0$ | $C_1$ |
| $\left\|\Phi^-\right\rangle_{AB}$ | $C_3$ | $C_2$ | $C_1$ | $C_0$ |

Now, the author would like to describe the quantum dialogue protocol in detail. Similar to Deng *et al.*'s two-step QSDC protocol[3], the present protocol transmits two atom sequences in a two-step manner, and encodes secret messages with unitary operations. Moreover, two security checks are also adopted to ensure its transmission security. The present protocol is composed of the following steps.

**Step 1: Preparation of initial quantum states.** Alice prepares $2N$ Bell states $\left\{(A_1, B_1), (A_2, B_2), \cdots, (A_{2N}, B_{2N})\right\}$, where each two adjacent Bell states $(A_{2n-1}, B_{2n-1})$ and $(A_{2n}, B_{2n})$ $(n = 1, 2, \cdots, N)$ are in the same state. The atom $A$ $(B)$ from each Bell state is picked out to form an ordered atom sequence $S_A$ $(S_B)$. That is, $S_A = \left\{A_1, A_2, \cdots, A_{2N}\right\}$ and $S_B = \left\{B_1, B_2, \cdots, B_{2N}\right\}$. For the first security check, Alice prepares another batch of sample Bell states and randomly inserts atom $A$ $(B)$ into sequence $S_A$ $(S_B)$. Consequently, sequence $S_A$ $(S_B)$ turns into a new sequence $S_A^{'}$ $(S_B^{'})$. Afterward, Alice sends $S_B^{'}$ to Bob, and keeps $S_A^{'}$ in her hand.

**Step 2: The first security check.** As soon as Bob announces Alice his receipt of $S_B^{'}$, they begin to implement the first security check. Alice tells Bob the positions of the sample atom $B$ in $S_B^{'}$. Afterward, Bob randomly chooses $Z$ -basis $\left(\{|g\rangle,|e\rangle\}\right)$ or $X$ -basis $\left(\{|+\rangle,|-\rangle\}\right)$ to measure the sample atom $B$ in $S_B^{'}$, and informs Alice of his measurement basis and measurement results. Alice chooses the same measurement basis as Bob's to measure the sample atom $A$ in $S_A^{'}$. Comparing her own measurement results with Bob's, Alice can judge whether there exists eavesdropping or not. If the channel is secure, their measurement results should have deterministic correlations shown in Eqs.(1)-(4) , and the communication will be continued. Otherwise, the error rate will go beyond the threshold so that the communication is given up.

**Step 3: Alice's encoding.** After getting rid of the sample atoms, $S_A^{'}$ and $S_B^{'}$ turn back into $S_A$ and $S_B$, respectively. Both Alice and Bob divide their own sequence into groups, each of which contains two adjacent atoms. As a result, $\left(A_{2n-1}, A_{2n}\right)$ and $\left(B_{2n-1}, B_{2n}\right)$ form a group in $S_A$ and $S_B$, respectively $\left(n=1,2,\cdots,N\right)$. Then, Alice imposes a unitary operation $U_{i_n j_n}^{A}$ on $A_{2n-1}$ to encode her two-bit secret message $\left(i_n, j_n\right)$ $\left(n=1,2,\cdots,N\right)$. Afterward, for the second security check, Alice prepares a large number of sample single atoms, which are randomly in one of the four states $\left\{|g\rangle,|e\rangle,|+\rangle,|-\rangle\right\}$, and randomly inserts them into $S_A$. Consequently, $S_A$ turns into a new sequence $S_A^{''}$. Afterward, Alice sends $S_A^{''}$ to Bob.

**Step 4: The second security check.** As soon as Bob confirms Alice his receipt of $S_A^{''}$, Alice tells Bob the positions and the preparation basis of sample single atoms in $S_A^{''}$. Afterward, Bob measures sample single atoms in the same basis as Alice told and informs Alice of his measurement results. Comparing the initial states of sample single atoms with Bob's measurement results, Alice can judge whether there exists eavesdropping or not. If the channel is secure, the communication will be continued. Otherwise, the error rate will go beyond the threshold so that the communication is given up.

**Step 5: Bob's encoding.** After getting rid of the sample atoms, $S_A^{''}$ turns back into $S_A$ again. Bob picks up one atom from each sequence of $S_A$ and $S_B$ in order, and stores each two adjacent Bell states as a group. As a result, the $n$ th group contains two Bell states $\left\{\left(U_{i_n j_n}^{A} A_{2n-1}, B_{2n-1}\right),\left(A_{2n}, B_{2n}\right)\right\}$. In order to know the initial state of the $n$ th group, Bob imposes the Bell-basis measurement on $\left(A_{2n}, B_{2n}\right)$. According to his Bell-basis measurement result, Bob reproduces a new $\left(A_{2n}, B_{2n}\right)$ on which no measurement was imposed. Afterward, Bob imposes a unitary operation $U_{k_n l_n}^{B}$ on the new atom $B_{2n}$ to encode his two-bit secret message $\left(k_n, l_n\right)$ $\left(n=1,2,\cdots,N\right)$. Accordingly, the $n$ th group changes from $\left\{\left(U_{i_n j_n}^{A} A_{2n-1}, B_{2n-1}\right),\left(A_{2n}, B_{2n}\right)\right\}$ to $\left\{\left(U_{i_n j_n}^{A} A_{2n-1}, B_{2n-1}\right),\left(A_{2n}, U_{k_n l_n}^{B} B_{2n}\right)\right\}$.

**Step 6: Quantum dialogue.** Bob simultaneously sends the atoms $U_{i_n j_n}^{A} A_{2n-1}$ and $A_{2n}$ into one single-mode cavity so that they simultaneously interact with it driven by a classical field. In the meanwhile, Bob simultaneously sends the atoms $B_{2n-1}$ and $U_{k_n l_n}^{B} B_{2n}$ into the other single-mode cavity so that they also simultaneously interact with it driven by a classical field. The interaction time and Rabi frequency are ensured to satisfy $\lambda t=\pi/4$ and $\Omega t=\pi$ in both cases by Bob. Then, Bob measures the states of the atoms $U_{i_n j_n}^{A} A_{2n-1}$ and $A_{2n}$ and the states of the atoms $B_{2n-1}$ and $U_{k_n l_n}^{B} B_{2n}$ under Z-basis, respectively, after they fly out their own single-mode cavity. Bob can infer out which result collection $\left\{U_{i_n j_n}^{A} A_{2n-1} A_{2n}, B_{2n-1} U_{k_n l_n}^{B} B_{2n}\right\}$ belongs to from Eqs.(9)-(12). As a result, Bob can know the four possible kinds of states about $\left\{\left(U_{i_n j_n}^{A} A_{2n-1}, B_{2n-1}\right),\left(A_{2n}, U_{k_n l_n}^{B} B_{2n}\right)\right\}$ from Table 1. With $U_{k_n l_n}^{B}$ and the knowledge of the initial state from his Bell-basis measurement on $\left(A_{2n}, B_{2n}\right)$, Bob can read out Alice's two-bit secret message $\left(i_n, j_n\right)$ from Table 1. On the other hand, in order to realize quantum dialogue, Bob informs Alice of the result collection through the classical channel. As a result, Alice can also know the four possible kinds of states about $\left\{\left(U_{i_n j_n}^{A} A_{2n-1}, B_{2n-1}\right),\left(A_{2n}, U_{k_n l_n}^{B} B_{2n}\right)\right\}$ from Table 1. With $U_{i_n j_n}^{A}$ and the initial state she prepared, Alice can infer out Bob's two-bit secret message $\left(k_n, l_n\right)$ from Table 1.

So far, the description of a quantum dialogue protocol via cavity QED has been finished. For clarity, take $\left(A_1, B_1\right)$ and $\left(A_2, B_2\right)$ as example to further illustrate the bidirectional communication process. Suppose that the first two-bit secret messages of Alice and Bob are **01** and **10**, respectively, and that both $\left(A_1, B_1\right)$ and $\left(A_2, B_2\right)$ are in the state of $\left|\Psi^-\right\rangle$. The parameters for cavity QED are chosen to satisfy that $\lambda t = \pi / 4$ and $\Omega t = \pi$ . As a result, $(A_1, B_1)$ and $(A_2, B_2)$ will evolve in the following way:

$$\left.\begin{array}{l} \left|\Psi^-\right\rangle_{A_1B_1} \Rightarrow \sigma_x^A \otimes \left|\Psi^-\right\rangle_{A_1B_1} = \left|\Phi^-\right\rangle_{A_1B_1} \\ \left|\Psi^-\right\rangle_{A_2B_2} \Rightarrow i\sigma_y^B \otimes \left|\Psi^-\right\rangle_{A_2B_2} = \left|\Phi^+\right\rangle_{A_2B_2} \end{array}\right\} \Rightarrow \left.\begin{array}{l} |gg\rangle_{A_1A_2} |gg\rangle_{B_1B_2} \\ |ee\rangle_{A_1A_2} |ee\rangle_{B_1B_2} \\ |ge\rangle_{A_1A_2} |ge\rangle_{B_1B_2} \\ |eg\rangle_{A_1A_2} |eg\rangle_{B_1B_2} \end{array}\right\} \rightarrow \mathbb{C}_1 \tag{13}$$

With $\mathbb{C}_1$ and $i\sigma_y^B$ , Bob can read out from Table 1 that Alice's first two-bit secret message is **01**, since he knows the prepared initial state from the Bell-basis measurement on $(A_2, B_2)$ . On the other hand, Bob informs Alice of $\mathbb{C}_1$ through the classical channel. With $\mathbb{C}_1$ and $\sigma_x^A$ , Alice can read out from Table 1 that Bob's first two-bit secret message is **10,** since she prepares the initial state by herself.

## 3 Security analysis

Firstly, the general active attacks from an outside eavesdropper, Eve, is considered. Apparently, the present protocol sends two atom sequences form Alice to Bob in a two-step manner, similar to Deng *et al*'s two-step QSDC[3]. Actually, during the transmission of $S_A^{''}$ , Eve can only disturb it but steal none of secret messages, as no one can distinguish a Bell state with just one atom. Why this happens is due to the state property of atom $A$ . That is, atom $A$ is a complete mixed state, having a reduced density matrix equal to $\frac{1}{2}I$ . As a result, the security of the present protocol only relies on the transmission of $S_B^{'}$ . Apparently, the deterministic entanglement correlation between two atoms from a sample Bell state is used by the first security check to detect whether there exists eavesdropping during this transmission. Its validity against the general active attacks, such as the intercept-resend attack, the measure-resend attack and the entangle-measure attack, is now demonstrated in detail. (I)The intercept-resend attack. Eve intercepts $S_B^{'}$ and sends her prepared fake sequence instead of $S_B^{'}$ to Bob. Since the original entanglement correlation between atom $A$ and atom $B$ has been disturbed, Eve can be detected with probability $50\%$ .[26,28,29,34,40] (II) The measure-resend attack. After intercepting $S_B^{'}$ , Eve measures it and resends it to Bob. Since the measurement basis that Alice and Bob choose are not always consistent with those of Eve, Eve can be detected with probability $25\%$ .[26,28,29,34,40] (III) The entangle-measure attack. Eve may steal partial information by entangling her auxiliary atom $|\varepsilon\rangle$ with the atoms in $S_B^{'}$ . Then the whole system will evolve into

$$\hat{E}|g\rangle|\varepsilon\rangle = \alpha_1 |g\rangle|\varepsilon_{00}\rangle + \beta_1 |e\rangle|\varepsilon_{01}\rangle ,$$
$$\hat{E}|e\rangle|\varepsilon\rangle = \beta_2 |g\rangle|\varepsilon_{10}\rangle + \alpha_2 |e\rangle|\varepsilon_{11}\rangle . \tag{14}$$

Apparently, Eve can be detected with probability $\zeta = |\beta_1|^2 = |\beta_2|^2$ when the security check is implemented under $Z$ -basis. [26,28,29,34,40]

Besides, whether the information leakage problem exists should also be analyzed. Still take $(A_1, B_1)$ and $(A_2, B_2)$ as example. In the present protocol, $(A_2, B_2)$ is the shared secret Bell state between two communicators so that Eve has no access to $(A_2, B_2)$ . As a result, Bob's announcement of $\mathbb{C}_1$ contains sixteen kinds of unitary operation combinations from two communicators, including $-\sum_{i=1}^{16} p_i \log_2 p_i = -16 \times \frac{1}{16} \log_2 \frac{1}{16} = 4$ bit information for Eve. Consequently, none of two communicators' secret messages leak out to Eve. Apparently, $(A_2, B_2)$ makes Bob aware of the prepared initial state, thus helps overcome the information leakage problem.

## 4 Discussions and conclusions

Obviously, the quantum dialogue protocol in Ref.[33] uses the entanglement swapping between any two initial Bell states together with the shared secret Bell state to solve the information leakage problem. As to the present protocol, the information leakage problem is solved by using the entanglement swapping between any two initial Bell states via cavity QED together with the shared secret Bell state. Therefore, it can be concluded that the present protocol can be regarded as the generalization of the quantum dialogue protocol in Ref.[33] in the case of cavity QED. In this way, it is not surprised that several implementation steps for them are similar.

Besides, the author would like to compare the present protocol with the protocols in Refs.[21,32] as all of them belong to the kind of quantum dialogue protocol via cavity QED. With regard to each round communication, the comparisons are summarized in Table 2, concentrating on four aspects, i.e., quantum resource, information-theoretical efficiency, quantum measurement, and information leakage. The information-theoretical efficiency defined by Cabello[41] is $\eta = b_s / (q_t + b_t)$, where $b_s$, $q_t$ and $b_t$ are the expected secret bits received, the qubits used and the classical bits exchanged between Alice and Bob.

Table 2. Comparisons among three quantum dialogue protocols via cavity QED for each round communication

| | The protocol of Ref.[21] | The protocol of Ref.[32] | The present protocol |
|---|---|---|---|
| Quantum resource | Two Bell states | Two Bell states | Two Bell states |
| Information-theoretical efficiency | 40% | 66.7% | 66.7% |
| Quantum measurement | Four Z-basis measurements | Two Bell-basis measurements and two Z-basis measurements | One Bell-basis measurement and four Z-basis measurements |
| Information leakage | Yes | No | No |

From Table 2, it is obvious that the protocol of Ref.[21] is meaningless, since partial of information leaks out to Eve. Moreover, the present protocol exceeds the protocol of Ref.[32] in quantum measurement as the Z-basis measurement is much easier to implement than the Bell-basis measurement in practice.

In summary, in this paper, a secure quantum dialogue protocol via cavity QED is suggested by using the evolution law of atom in cavity QED. The present protocol employs the two-step transmission and the unitary operation encoding. Two security checks are adopted to ensure its transmission security against the active attacks from an outside eavesdropper. Moreover, the present protocol avoids the information leakage problem by using the entanglement swapping between any two Bell states via cavity QED together with the shared secret Bell state. Compared with the previous information leakage resistant quantum dialogue protocol via cavity QED, the present protocol takes advantage in quantum measurement.

**Acknowledgements**

Funding by the National Natural Science Foundation of China (Grant No.11375152), and the Natural Science Foundation of Zhejiang Province (Grant No. LQ12F02012) is gratefully acknowledged.